\documentclass[referee]{aa} 
\usepackage{graphicx}
\usepackage{graphics}
\usepackage{colortbl}
\usepackage{txfonts}
\usepackage{natbib}
\newcommand{\HESS}{H.E.S.S.\,}
\newcommand{\this}{HESS\,J1813--178}

\newcommand{\dg}{\ensuremath{^\circ}}

\begin{document}
  \title{{\emph{XMM-Newton}} observations of \this\ reveal a composite
  Supernova remnant}

  \author{S.~Funk \inst{1,2} 
    \and J.~A.~Hinton~\inst{3} 
    \and Y.~Moriguchi~\inst{1,4}
    \and F.~A.~Aharonian~\inst{1} 
    \and Y.~Fukui~\inst{4}
    \and W.~Hofmann~\inst{1} 
    \and D.~Horns \inst{5}
    \and G.~P\"uhlhofer~\inst{6} 
    \and O.~Reimer~\inst{7} 
    \and G.~Rowell~\inst{8}
    \and R.~Terrier~\inst{9}
    \and J.~Vink~\inst{10}
    \and S.J.~Wagner\inst{6} 
  }
  
  \offprints{Stefan Funk (Stefan.Funk@slac.stanford.edu)}

  \institute{
    Max-Planck-Institut f\"ur Kernphysik, P.O. Box 103980, D 69029
    Heidelberg, Germany
    \and 
    Kavli Institute for Astroparticle Physics and Cosmology, SLAC, P.O
    Box 0029, CA-94025, USA 
    \and
    School of Physics \& Astronomy, University of Leeds, Leeds LS2 9JT, UK
    \and 
    Department of Astrophysics, Nagoya University, Chikusa-ku,
    Nagoya 464-8602, Japan
    \and 
    Institut f\"ur Astronomie und Astrophysik, Universit\"at T\"ubingen, 
    Sand 1, D 72076 T\"ubingen, Germany
    \and
    Landessternwarte, Universit\"at Heidelberg, K\"onigstuhl, D 69117
    Heidelberg, Germany
    \and 
    Stanford University, HEPL \& KIPAC, Stanford, CA 94305-4085,
    USA
    \and 
    School of Chemistry \& Physics, University of Adelaide,
    Adelaide 5005, Australia 
    \and 
    APC, 11 Place Marcelin Berthelot, F-75231 Paris Cedex 05, France
    \and 
    Astronomical Institute, University Utrecht, P.O. Box 80000,
    3508TA Utrecht, The Netherlands 
  }
  

  \abstract {}
      {We present X-ray and $^{12}$CO(J=1--0) observations of the
      very-high-energy (VHE) $\gamma$-ray source \object{\this}\ with
      the aim of understanding the origin of the
      $\gamma$-ray emission.} 
      { High-angular resolution X-ray studies of the VHE
      $\gamma$-ray emission region are performed using 18.6~ks of {\emph{XMM-Newton}}
      data, taken on \this\ in October 2005. Using this dataset we are able
      to undertake spectral and morphological studies of the X-ray emission
      from this object with greater precision than previous studies. NANTEN $^{12}$CO(J=1--0) data
      are used to search for correlations of the $\gamma$-ray emission
      with molecular clouds which could act as target material for
      $\gamma$-ray production in a hadronic scenario.} 
      {The NANTEN $^{12}$CO(J=1--0) observations show a giant
      molecular cloud of mass $2.5 \times 10^5$ M$_{\sun}$ at a
      distance of 4~kpc in the vicinity of \this. Even though there is 
      no direct positional coincidence, this giant cloud might have
      influenced the evolution of the $\gamma$-ray source and its
      surroundings. The X-ray data show a highly absorbed
      (n$_{\mathrm{H}} \sim 1. \times 10^{23}$ cm$^{-2}$) non-thermal
      X-ray emitting object coincident with the previously known ASCA 
      source \object{AX\,J1813--178} showing a compact core and an extended tail
      towards the north-east, located in the centre of the radio
      shell-type Supernova remnant (SNR) \object{G12.82--0.2}. This central
      object shows morphological and spectral resemblance to a Pulsar Wind 
      Nebula (PWN) and we therefore consider that this object is very
      likely to be a \emph{composite SNR}. We discuss the
      scenario in which the $\gamma$-rays originate in 
      the shell of the SNR and the one in which they originate in the
      central object in terms of a time-dependent one-zone
      leptonic model. We demonstrate, that in order to connect the
      core X-ray emission to the VHE $\gamma$-ray emission electrons
      have to be accelerated to energies of at least 1~PeV. We 
      conclude that if indeed the X-rays are connected to the
      VHE $\gamma$-rays HESS\,J1813--178 has to be a Galactic
      {\emph{Pevatron}}.   
}
      {}
  \authorrunning{S. Funk et al.}  
  \titlerunning{{\emph{XMM-Newton}} and NANTEN observations of \this}
  \keywords{ISM: Supernova remnants, plerions -- ISM: individual
      objects: \this, G12.82--0.02, AX\,J1813--178 -- X-rays:
      observations -- gamma-rays: observations
  }  
  \maketitle


\section{Introduction}
During a survey of the Galactic plane in very high-energy (VHE)
$\gamma$-rays the atmospheric Cherenkov telescope system \HESS\ found
more than a dozen new $>$100 GeV $\gamma$-ray sources~\citep{HESSScan,
HESSScanII}. Even though some of the new $\gamma$-ray sources can be
identified with counterparts at other wavebands, for example the
\emph{microquasar} \object{LS\,5039}~\citep{HESSLS5039, HESSLS5039II},
most of them lack a compelling positional
counterpart~\citep{FunkBarcelona}. A firm counterpart identification
for a $\gamma$-source at other wavebands requires not only a spatial
association, but also a viable $\gamma$-ray emission mechanism and a
consistent multi-frequency picture. X-ray observations with
instruments such as {\emph{XMM-Newton}} and {\emph{Chandra}} combine
high-angular-resolution and high-sensitivity observations with the
possibility to trace relativistic electrons via their synchrotron
emission. These instruments are thus ideally suited to the
identification of VHE $\gamma$-ray objects. This article reports on
multi-frequency observations of \this\ with the {\emph{XMM-Newton}}
X-ray satellite. Dense molecular regions that could act as target
material for hadronic interactions have been studied with the NANTEN
CO-telescope in the direction of \this\ to test for an origin of the
$\gamma$-ray emission in the decay of neutral pions produced in
proton-proton collisions.

\this\ was in fact the first source discovered during the survey. It
was subsequently re-observed to increase the statistical significance
and the signal in 9.7 hours of live-time comprised 340 $\gamma$-ray
events at a significance of $\sim 14$ standard deviations. While many
of the new sources found in the survey showed a rather large source
extension, \this\ has a compact nature and is only slightly extended
(Gaussian width $\sigma = 2.2\arcmin \pm 0.4\arcmin$) with respect to
the point-spread function (PSF) of the \HESS\ instrument. The flux of
the source is $(14.2 \pm 1.1_{\mathrm{stat}}) \times 10^{-12}$ photons
cm$^{-2}$ s$^{-1}$ above 200~GeV and the energy spectrum follows a
power-law with photon index $2.09 \pm 0.08$. \this\ is located in the
Galactic plane at $18^{\mathrm{h}}13^{\mathrm{m}}38^{\mathrm{s}}$,
$-17^{\mathrm{d}}50\arcmin33\arcsec$. \this\ was subsequently also
detected by the MAGIC Cherenkov telescope confirming the basic
properties of the source~\citep{1813MAGIC}.

At the time of the first publication of this source~\citep{HESSScan},
\this\ was still considered as unidentified. However, subsequently to
the \HESS\ publication, the source was reported to show compelling
positional coincidence with several sources in other energy
bands~\citep{Brogan,Helfand, Ubertini}. \this\ coincides with the
previously unpublished archival ASCA source AX\,J1813--178 in the
2--10~keV band~\citep{Brogan}. This unresolved ASCA source with a
rather hard photon index of $1.8\pm0.4$ and a flux of $7 \times
10^{-12}$ erg~cm$^{-2}$~s$^{-1}$ (2--10 keV) was one of the relatively
bright sources detected in the ASCA Galactic Plane
survey~\citep{ASCA}.  AX\,J1813--178 exhibits a highly absorbed
(n$_{\mathrm{H}} = (1.1 \pm 0.2) \times 10^{23}$ cm$^{-2}$)
non-thermal spectrum without indications of emission lines. The value
of n$_{\mathrm{H}}$ is significantly larger than the total column
density through the Galaxy in this direction, suggesting that the
source is embedded in a dense environment. Positionally in coincidence
is the unresolved hard X-ray source \object{IGR\,J18135--1751}
detected in the INTEGRAL Galactic plane survey in the 20--100 keV
energy band~\citep{Ubertini}. The source connects to the energy
spectrum of the ASCA source, showing however a softer energy spectrum,
suggesting a spectral break in the X-ray spectrum around
10~keV. Recently the source has also been observed with the
{\emph{Swift}} X-Ray Telescope (XRT) that found an X-ray source
coincident with the ASCA source. The XRT data confirm the hard
spectral index and the high column density towards the source present
in the ASCA data~\citep{Swift1813}.

The region surrounding \this\ was also covered by observations in the
20~cm and 90~cm band by the VLA. These data showed a faint non-thermal
radio source (G12.82--0.02) visible at the position of
\this~\citep{Brogan, White}. G12.82--0.02 lies in the projected
vicinity (distance $\sim15$\arcmin) of the \object{W\,33} region. This
region is known to contain ultra-compact HII
regions~\citep{UltracompactHII} and contains methanol, hydroxyl, and
water masers and other tracers of recent star formation. While \HESS\,
ASCA, and INTEGRAL lack the spatial resolution to resolve the object,
the angular resolution of the VLA allows a shell-like radio structure
of diameter 2.5\arcmin to be resolved. The flux at 20~cm was reported
as $0.65 \pm 0.1$ Jy, the flux at 90~cm was given as $1.2 \pm 0.08$
Jy, yielding a non-thermal radio spectrum of index $\alpha = -0.42 \pm
0.03$ ($j_\nu \propto \nu^{\alpha}$). {\emph{Spitzer}} Space telescope
data from the GLIMPSE survey at 8~$\mu$m~\citep{Spitzer} show no signs
of dust emission in positional coincidence with G12.82--0.02, a
finding that supports the non-thermal nature of the source. The radio
spectrum, morphology and the lack of coinciding IR-Emission
led~\citet{Brogan} to the conclusion that the radio structure is a
previously unknown young shell-type Supernova remnant (SNR). In the
Parkes multi-beam pulsar survey (PMBS) no pulsar close to G12.82--0.02
has been found thereby determining the 1.4~GHz flux density upper
limit at a rather constraining level of
0.2~mJy~\citep{Parkes}. Currently the only constraints on the distance
of the object can be derived from the strong absorption found in the
X-ray data that suggests that AX\,J1813--178 is located behind W~33 at
a distance larger than 4~kpc.

Assuming an association between \this\ and the radio source
G12.82--0.02, the question remains whether the thus far unresolved
X-ray and $\gamma$-ray emission originates from the shell of the SNR,
or rather from a Pulsar Wind Nebula (PWN) embedded within the
shell.~\citet{Brogan} concluded from the fact that the radio spectrum
nearly directly extrapolates into the ASCA X-ray spectrum, that both
the radio shell and the ASCA source should indeed be the same source
and therefore that all the emission was connected to the radio shell.

This article reports on high-angular resolution X-ray observations
performed with {\emph{XMM-Newton}} with the goal of pinning down the
origin of the high-energy X-ray and $\gamma$-ray emission. Another
option for the $\gamma$-ray emission is the interaction of accelerated
hadrons with dense molecular material.  The bright ultra-compact HII
region W\,33~\citep{UltracompactHII} is the nearest star forming
region, located $\sim 10\arcmin$ away from \this. Dense gas in the
W\,33 region could act as target material for the VHE $\gamma$-ray
generation in hadronic interactions, therefore a study of the
$^{12}$CO(J=1--0) distribution performed during the NANTEN survey of
the Galactic plane~\citep{NantenCO} towards \this\ is included.

\section{{\emph{XMM-Newton}} Observations of the region}
\begin{figure*}
  \centering
  \includegraphics[width=0.85\textwidth]{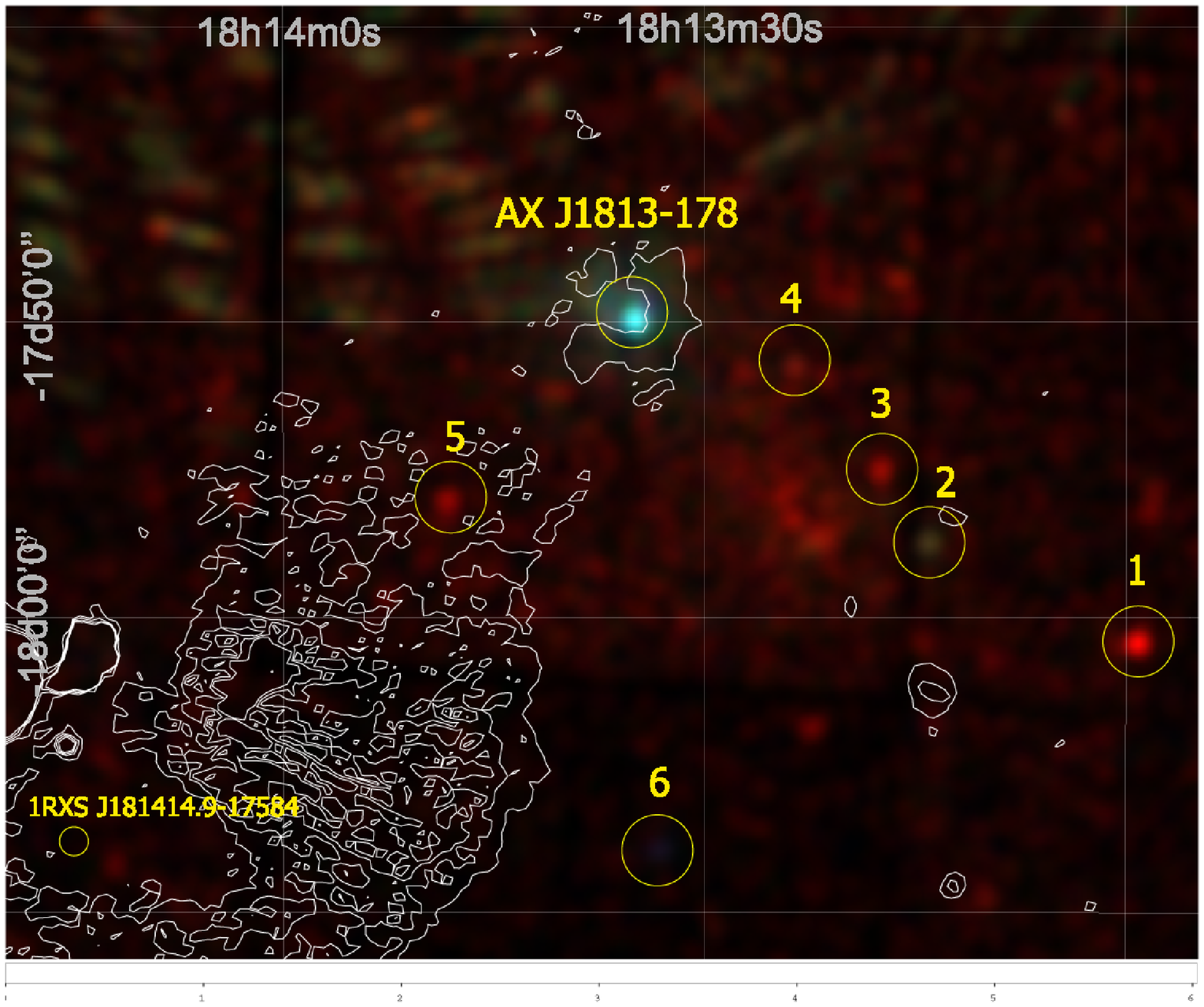}
  \includegraphics[width=0.75\textwidth]{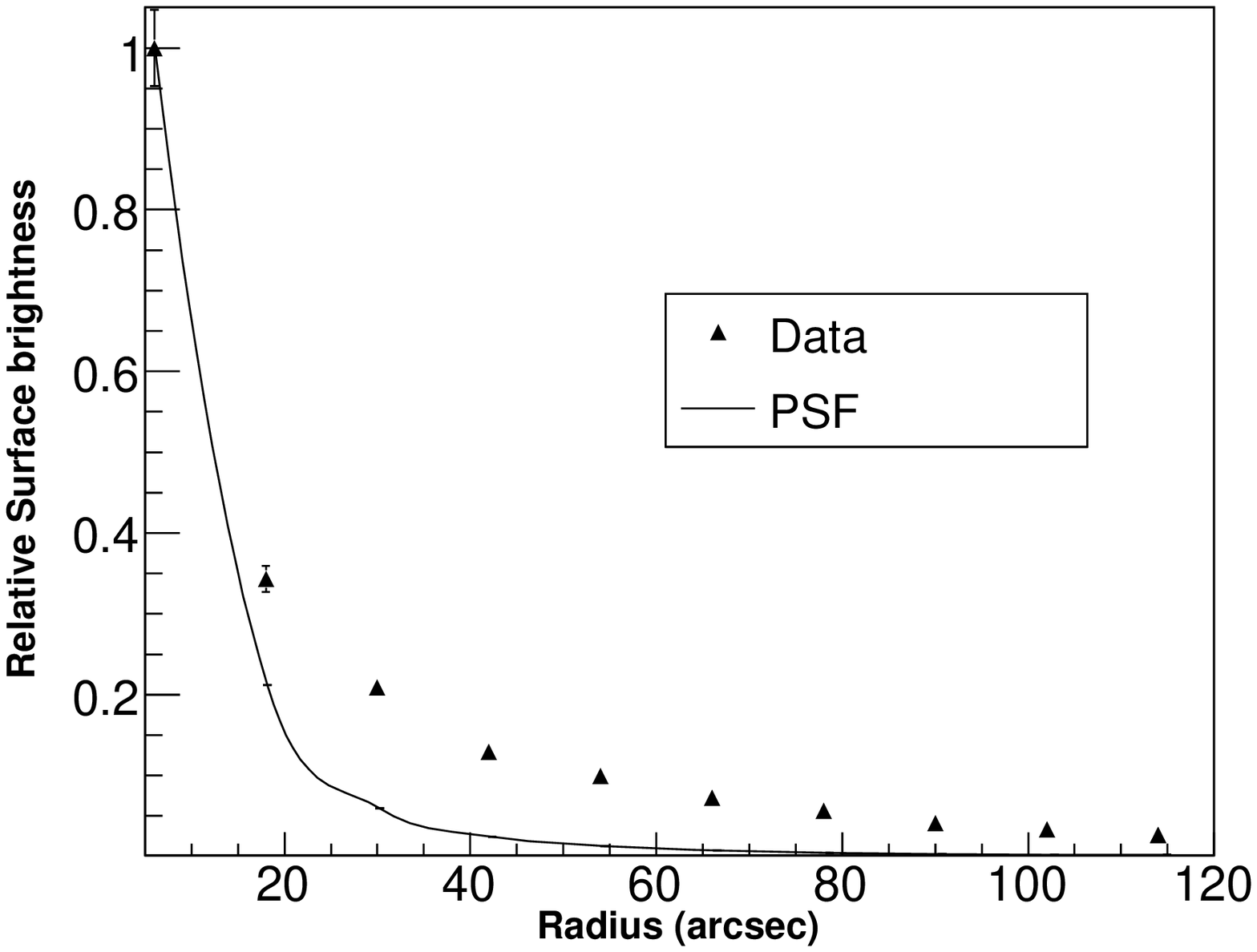}
  \caption{{\bf{Top:}} A composite false-colour image of the EMOS1 and
  EMOS2 count maps, slightly smoothed with a Gaussian kernel of width
  3 image pixels. Different colours correspond to 0.5~keV--2~keV
  (red), 2~keV--4.5~keV (green), and 4.5~keV--10~keV
  (blue). AX\,J1813--178 appears as a bright, hard-spectrum source
  with a tail towards the north-west. The north-western edge of the
  field of view shows some soft stray-light contamination, caused by
  the low mass X-ray binary \object{GX\,13+1}. Also shown are several
  other X-ray sources that were found with {\emph{emldetect}} at
  energies above 4.5~keV. These are typically fainter and have a
  softer photon index. The white contours denote the VLA 20~cm radio
  emission as already reported by~\citet{Brogan}. The shell-like radio
  structure surrounding AX\,J1813--178 is clearly
  visible. {\bf{Bottom:}} Radial profile of the central source
  AX\,J1813--178 in comparison to the PSF for this data set.}
  \label{fig::XRayFoV}
\end{figure*}

\this\ was observed with {\emph{XMM-Newton}} on the 14th of October
2005 for 18.6~ksec. The instrument cameras (EPIC MOS1, MOS2, PN) were
operated in full-frame mode with a medium filter to screen out optical
and UV light. The data were processed with the {\emph{XMM-Newton}}
Science Analysis Software (SAS) version 7.0 as well as with the
Extended Source Analysis Software package (XMM-ESAS) version
1.0. Standard data reduction and calibration procedures were applied
to the data and the total data set amounts to 13.6~ksec after the
observations have been cleaned of flaring background caused by soft
protons. Figure~\ref{fig::XRayFoV} shows the resulting combined count
map of the EMOS1 and EMOS2 detectors for three different energy bands
(red: 0.5~keV to 2~keV, green: 2~keV to 4.5~keV, blue: 4.5~keV to
10~keV) smoothed with a Gaussian kernel of 3 pixels. No shell-like
emission as seen in the VLA data but rather a compact X-ray source
with an extended tail towards the north-west is visible at the centre
of the EMOS-Cameras. This source is in positional coincidence with the
previously reported ASCA source AX\,J1813--178. There is still a
contamination of soft photons in the dataset. This had been seen
before in the ASCA data and can be attributed to the bright, close-by
($\sim 0.7^{\circ}$) low mass X-ray binary GX\,13+1. To investigate
the influence of the soft photon contamination in the background
regions for the spectral studies, three separate methods were applied
to estimate the background as discussed later in the text. The results
for the different methods show good agreement, enhancing the
confidence that the background level in the spectral studies was
estimated correctly. There is however still the possibility of soft
photon contamination in the source spectrum, especially at energies
below 2~keV, since the source appears strongly absorbed below this
energy.

\begin{table*}[h]
    \caption{X-ray sources other than AX\,J1813--178 detected in the
    field of view surrounding HESS\,J1813--178 using the detection
    algorithm \emph{emldetect}. The second column gives the name
    recommended by the {\emph{XMM-Newton}} SOC and the IAU for
    serendipitous {\emph{XMM-Newton}} source detections. Column 3 and
    4 give J2000 coordinates with a typical error on the position of
    1\arcsec. Column 5 gives the number of counts in a circle of
    radius 30\arcsec above 2~keV.}
    \label{tab::sourceDetection}
    \centering
    \begin{tabular}{|c| c | c c | c |}
      \hline
      Id & XMMUJ &  RA$_{2000}$ & Dec$_{2000}$ & Counts \\
      & & (h:m:s) & (d:\arcmin:\arcsec) & \\ \hline
      1 & 181259.2--175527 & 18:12:59.23 & --17:55:26.9 & 60\\
      2 & 181314.1--175344 & 18:13:14.11 & --17:53:44.1 & 238\\
      3 & 181318.0--175232 & 18:13:17.57 & --17:52:31.9 & 90\\
      4 & 181323.7--175041 & 18:13:23.68 & --17:50:41.2 & 138\\
      5 & 181348.3--175301 & 18:13:48.35 & --17:53:00.6  & 68\\ 	
      6 & 181333.3--175857 & 18:13:33.27 & --17:58:56.7 & 157\\ \hline
        & AX\,J1813--178   & 18:13:35.16 & --17:49:50.0 & 2192\\
      \hline
    \end{tabular}
\end{table*}

\begin{figure*}
  \centering
  \includegraphics[width=0.7\textwidth]{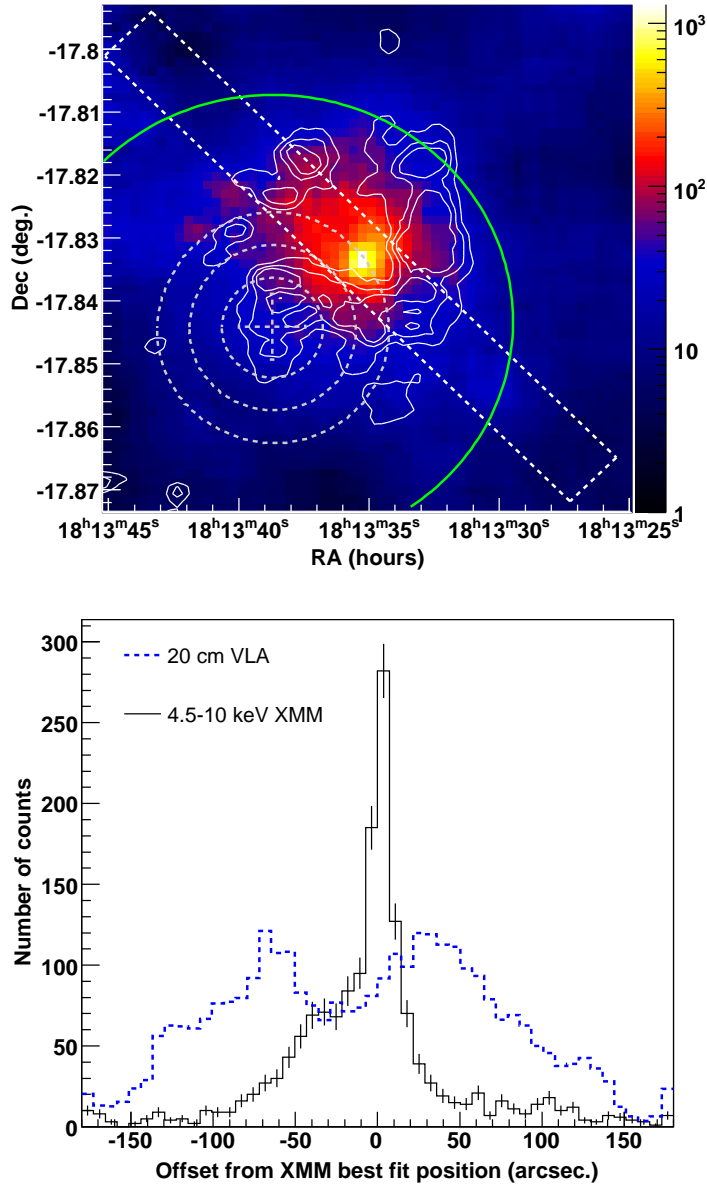}
  \caption{Comparison of radio and X-ray data of \this. {\bf Top:}
  {\emph{XMM-Newton}} counts map above 4.5~keV of the region
  surrounding \this\ (colour contours) smoothed with a Gaussian kernel
  of width 0.1\arcmin. The extended tail towards the north-east is
  visible in this figure. Overlaid is the 20~cm shell-like emission
  (white contours) as detected by the VLA~\citep{Brogan}. The
  difference in the images between the X-ray and the radio wavebands
  is apparent in this figure. Also shown are the positional contours
  (1, 2, 3 $\sigma$ error) of the best fit position of \this\ as given
  in~\citet{HESSScanII} (dashed circles) and the extension (solid
  green), covering both the radio and X-ray emitting region
  completely. {\bf Bottom:} Slice through the emission in radio and
  X-rays as plotted on the left hand side. The box in which the slices
  were determined is also given in the left panel (white box). The
  X-ray slice shows the compact core with the slice towards the
  north-east, whereas the radio slice shows the shell-like structure
  of the emission. }
  \label{fig::connection}
\end{figure*}

For the source position fitting the contamination was avoided by using
the source detection algorithm (\emph{emldetect}) as described
in~\citet{Snowden} for energies above 4.5~keV and for energies above
7.5~keV. Seven sources are detected in the band above 4.5~keV which
are not in the part of the field of view affected by the stray light
(see Table~\ref{tab::sourceDetection}), whereas above 7.5~keV only a
single X-ray source is detected. This source is extended and
coincident with AX\,J1813--178 at a best-fit position
18$^{\mathrm{h}}$13$^{\mathrm{m}}$35.16$^{\mathrm{s}}$,
--17$^{\mathrm{d}}$49\arcmin50.0\arcsec, with a statistical error on
the position of 2\arcsec.  This position is at a distance of $\sim
1\arcmin$ from the best fit position of \this, located well within the
extension of 2.2\arcmin\ of \this. Even though an a priori match of
the best fit X-ray and $\gamma$-ray positions is far from being
certain as e.g.\ shown in the PWN candidate HESS\,J1825--137 (in which
the \HESS\ position is significantly shifted from the peak of the
X-ray emission as shown in~\citet{HESSJ1825II}) this X-ray position is
compatible with the \HESS\ position, given a statistical error of
21\arcsec\ adding to a systematic error in the pointing accuracy of
the \HESS\ instrument of the similar size. The source position fitting
tool {\emph{emldetect}} also determines AX\,J1813--178 to be
incompatible with a point-source. Using a Gaussian model the extension
was determined to be $8.5\pm0.2$ image pixels, corresponding to an
projected width on the sky of $21\arcsec \pm 0.5\arcsec$. As apparent
in the slice through the source, shown in
Figure~\ref{fig::connection}, the Gaussian profile might not be the
correct representation of the source profile, which seems to have a
rather long tail towards the north-east. The source position
determined using the {\emph{XMM-Newton}} data is compatible with the
position of the only X-ray source found in the \emph{Swift} XRT
data~\citep{Swift1813} at
18$^{\mathrm{h}}$13$^{\mathrm{m}}$34.9$^{\mathrm{s}}$,
--17$^{\mathrm{d}}$49\arcmin53.2\arcsec with an error of
3.5\arcsec. The distance of the best fit position presented here to
the XRT position is 4.9\arcsec. The other X-ray sources within the
field of view listed in Table~\ref{tab::sourceDetection} are no known
X-ray sources.

\begin{figure*}
  \centering
  \includegraphics[width=0.8\textwidth, angle=270]{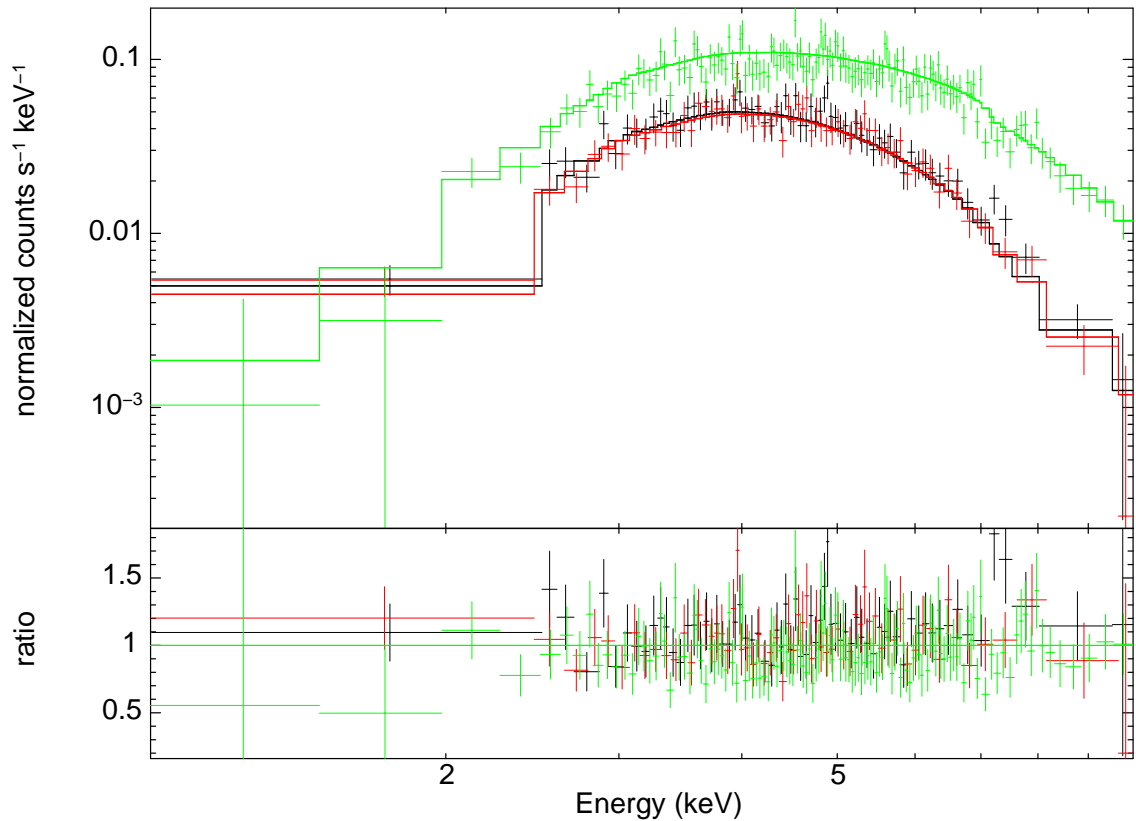}
  \caption{{\emph{XMM-Newton}} energy spectrum for all three detectors
  (EMOS1: black, EMOS2: red, EPN: green) for an extraction radius of
  75\arcsec, rebinned to yield a minimum significance of 5$\sigma$ in
  each bin, summing a maximum of 100 bins into a single one bin. The
  straight lines show the combined fit to these data (folded with the
  instrumental response which is different for the three
  detectors). The lower panel shows the residuals of the data to the
  fit. As can be seen, the power-law fit provides a reasonable
  description of the data. Also apparent is the absence of prominent
  line emission that lends support to the non-thermal nature of the
  emission.}
  \label{fig::Spectrum}
\end{figure*}

For the spectral analysis XSPEC (version 12.2.1) was used and three
different background methods as describe above were applied to
estimate the stray-light contamination on the spectrum of
AX\,J1813--178. The background was estimated a) from the south-western
part of the field of view, where the stray-light contamination is
apparently lower b) from a ring around the source with inner radius
100\arcsec and outer radius 200\arcsec and c) from blank field
observations as described in~\citet{Snowden}. All three methods yield
consistent results for the spectral fit, in the following the
background derived from a ring around the source within the same field
of view will be used. Different extraction radii of size 50\arcsec,
75\arcsec, and 100\arcsec\ were used to determine the spectrum of the
extended emission. All EMOS1, EMOS2, and EPN data were fit
simultaneously. In general the data are well described by a single
power-law and the results of the spectral fitting are summarised in
Table~\ref{table::SpectralFitting}.  Figure~\ref{fig::Spectrum} shows
the {\emph{XMM-Newton}} spectrum for the medium size extraction radius
of 75\arcsec. As can be seen, no prominent line emission can be found
and analysis shows that the shape is incompatible with a black-body
radiation spectrum, confirming the non-thermal nature of the X-ray
emission. The results for the different extraction radii are well
compatible with each other in terms of column density and spectral
index. The column density determined is $n_{\mathrm{H}} \sim 10^{23}$
cm$^{-2}$, significantly higher than the total Galactic column density
in this region of the Galactic plane, $1.6 \times 10^{22}$
cm$^{-2}$~\citep{DickeyLockman}. The photon index is determined to be
$\sim 1.8$, the flux between 2 and 10~keV for the 75\arcsec\
extraction radius amounts is $\sim 7 \times 10^{-12}$erg cm$^{-2}$
s$^{-1}$. Both these values are well compatible with previous
measurements reported from the analysis of ASCA data (photon index:
$1.8 \pm 0.4$, F$_{\mathrm{2-10~keV}} = 7.0 \times 10^{-12}$ erg
cm$^{-2}$ s$^{-1}$) by ~\citet{Brogan}.

\begin{table*}
    \caption{Spectral properties of AX\,J1813--178 as determined by
    the spectral fitting routine. EMOS1, EMOS2, and EPN data were
    simultaneously fit with a single absorbed power-law. The fit
    parameters are the absorption density n$_{\mathrm{H}}$, the photon
    index $\Gamma$, and the norm at 1~keV, which is expressed by the
    flux between 2~keV and 10~keV. Different spectral shapes such as a
    black-body spectrum were incompatible with the data (just to give
    an example: the $\chi^2$/d.o.f. for a blackbody spectrum for the
    extraction region of size 75\arcsec amounts to 5700. The errors
    given here correspond to the 90\% confidence levels.}
    \label{table::SpectralFitting}
    \centering
    \begin{tabular}{|c| c c c | c c c c|}
      \hline
      Radius & \multicolumn{3}{c}{Counts} & n$_{\mathrm{H}}$ & $\Gamma$ & F$_{\mathrm{2-10~keV}}$
      & $\chi^2$/d.o.f. \\
      (arcsec)& EMOS1 & EMOS2 & EPN & $10^{22}$ cm$^{-3}$&  & $10^{-12}$ erg cm$^{-2}$ s$^{-1}$ & \\ \hline
      \hline
      50 & 2004 & 1805 & 3930 & 12.0$^{+1.0}_{-0.7}$ & 1.81$^{+0.16}_{-0.11}$ & 5.2 & 3177/3001\\
      75 & 2957 & 2769 & 6194 & 10.6$^{+0.7}_{-0.7}$ & 1.72$^{+0.12}_{-0.12}$ & 6.8 & 3233/3001\\
     100 & 3794 & 3592 & 8397 & 10.6$^{+0.2}_{-0.3}$ & 1.88$^{+0.05}_{-0.03}$ & 6.9 & 3228/3001\\
      \hline 
    \end{tabular}
\end{table*}

The EPN data within a narrow region (5\arcsec) surrounding
AX\,J1813--178 was searched for periodic emission from any possible X-ray
pulsar. The timing resolution of the EPN detector in full-frame mode
is 73.2~ms. No significant periodicity was found.
Also no evidence for variability was detected in this dataset. Future
X-ray timing observations with {\emph{XMM-Newton}} or {\emph{Chandra}}
and future Multibeam radio observations might shed further light on
the existence possible underlying pulsar.

\section{NANTEN-Observations towards \this}

\begin{figure*}
  \centering
  \includegraphics[width=0.6\textwidth]{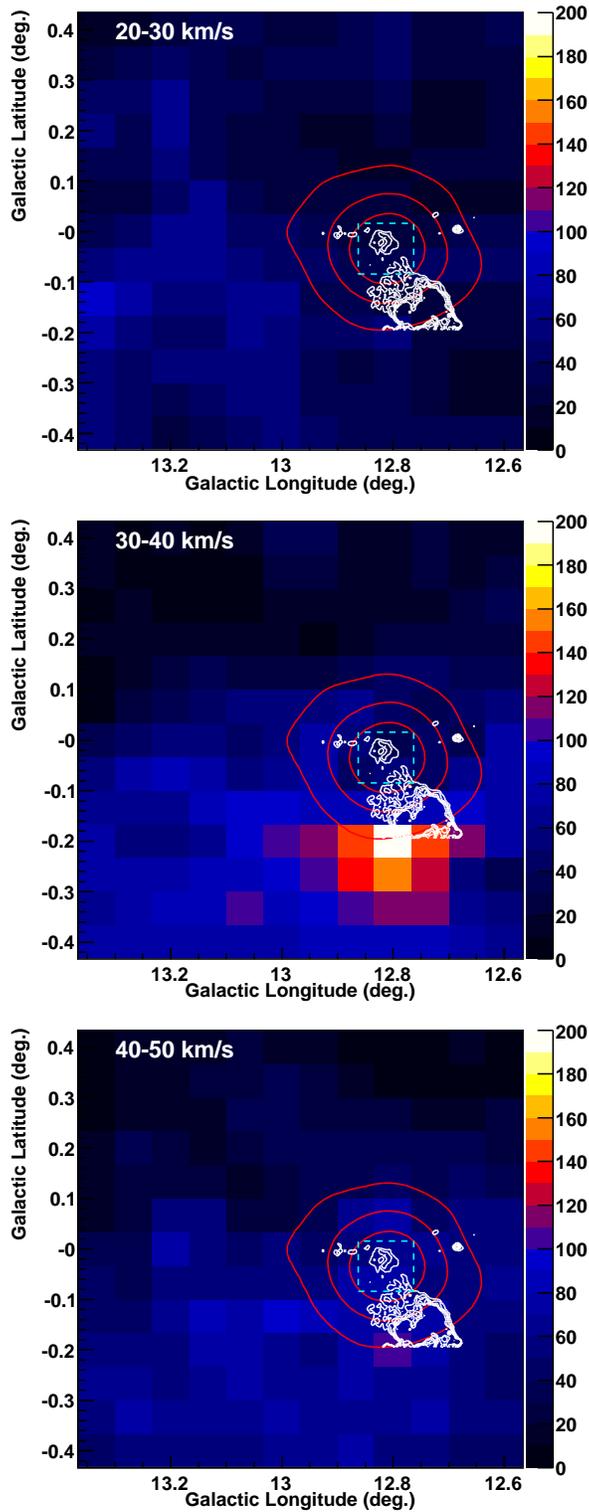}
  \caption{NANTEN $^{12}$CO(J=1--0) data towards the region
  surrounding \this\ for three different velocity ranges. Please note
  that this figure is in Galactic coordinates. The colour scale shows
  the CO emission in units of K~km~s$^{-1}$, the red contours the
  \HESS\ VHE $\gamma$-ray excess emission (smoothed with a Gaussian of
  0.05\dg\ width and plotted in equidistant excess contours), whereas
  the white contours show the 20~cm VLA contours~\citep{Brogan}. The
  dashed cyan box indicates the size of the area shown in
  Fig.~\ref{fig::connection}. The shell-type radio emission at the
  centre of this region is the SNR G12.82--0.02. As can be seen from
  this figure, the giant molecular cloud at 30--40~km/s (corresponding
  to $\sim 4$~kpc) is not in coincidence with \this, could however be
  associated with the star-forming region W\,33 to the south of
  \this.}
  \label{fig::Nanten}
\end{figure*}

\begin{figure*}
  \centering
  \includegraphics[width=0.9\textwidth]{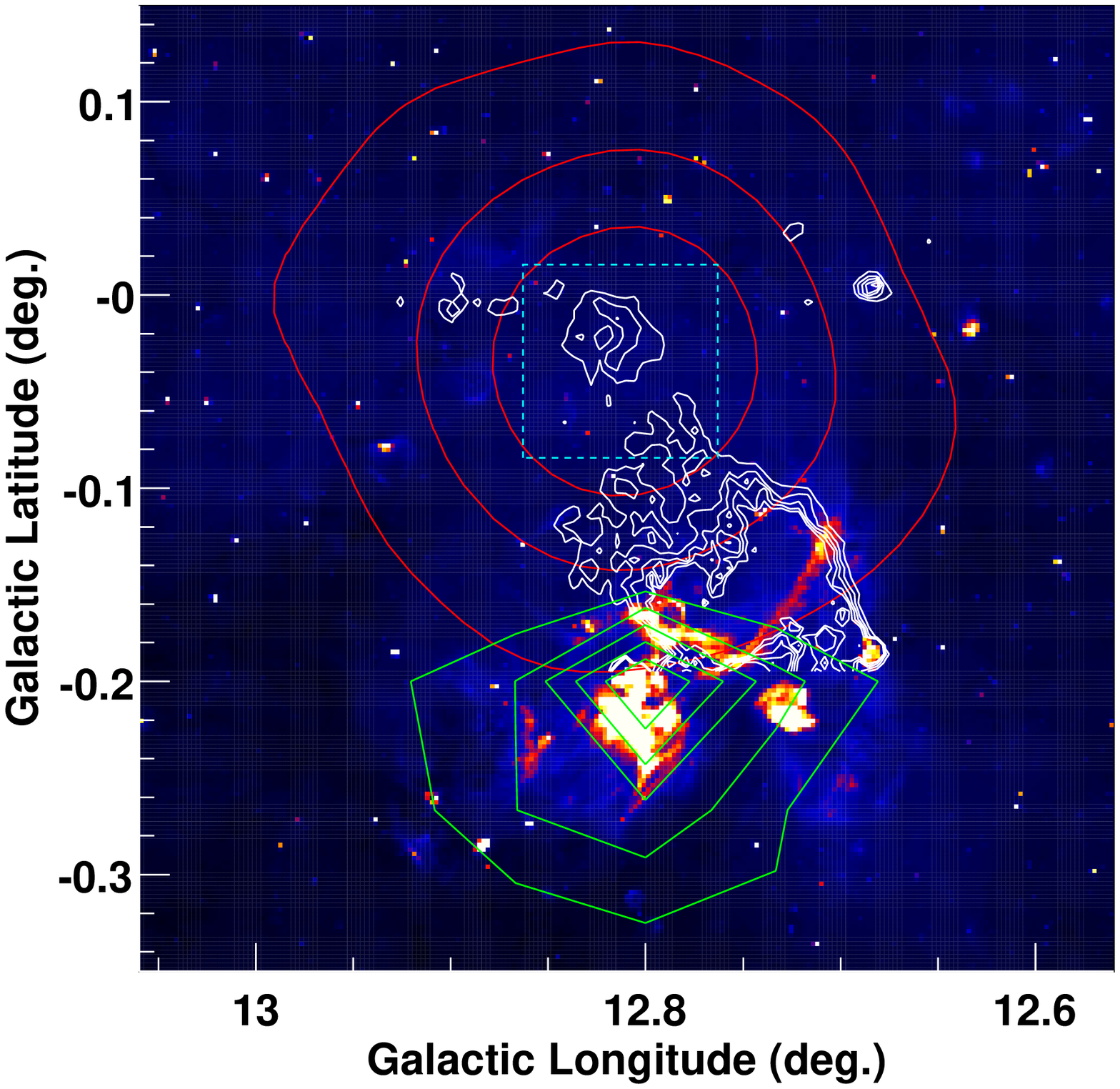}
  \caption{{\emph{Spitzer}} Space telescope data of the GLIMPSE survey
  at 8~$\mu$m~\citep{Spitzer} show in Galactic coordinates. The red
  contours show again the \HESS\ VHE $\gamma$-ray excess emission
  (smoothed with a Gaussian of 0.05\dg\ width and plotted in
  equidistant excess contours), whereas the white contours show the
  20~cm VLA contours~\citep{Brogan}. The green contours show the
  NANTEN CO-Emission contours at the distance of 4~kpc, showing the
  position of the GMC. The dashed cyan box indicates the size of the
  area shown in Fig.~\ref{fig::connection}. No dust emission is
  coincident with the VHE $\gamma$-ray source as reported already
  by~\citet{Brogan}, supporting the non-thermal nature of the
  source. The {\emph{Spitzer}} and NANTEN data are clearly correlated
  on this scale.}
  \label{fig::Spitzer}
\end{figure*}

The $^{12}$CO(J=1--0) observations were performed with the 4-meter
mm/sub-mm telescope, NANTEN, at Las Campanas Observatory of Carnegie
Institutions of Washington, Chile.  The NANTEN $^{12}$CO(J=1--0)
survey of the Galactic plane shows that the velocity of the $^{12}$CO
emission towards \this\ ranges from vLSR ({\emph{velocity local
standard of rest}}, i.e. the average velocity of stars in the solar
neighbourhood) $\sim 0$~km/s to $\sim60$~km/s. While the CO emission
at vLSR less than 10~km/s is likely due to local clouds, the emission
at velocities in excess of vLSR $\sim$ 20~km/s is likely from clouds
beyond 2 kpc.  The best known object in the region surrounding \this\
is W\,33, a complex of HII regions, a typical example of a massive
star forming region. The integrated intensity distribution at vLSR =
30--40~km/s shows an enhancement in close projected vicinity to \this\
as shown in Fig.~\ref{fig::Nanten}. To convert the velocity range into
a distance the model by~\citet{Brand} has been used and the rather
large uncertainty in kinematic distance for directions close to the
Galactic centre has to be taken into account. The W\,33 complex has
been studied fairly extensively in the past and its distance has been
estimated to be $\sim 4$~kpc (kinematic distance derived from
molecular line observations see e.g. \citep{Reifenstein, Goldsmith,
Mitchell}). The most massive molecular cloud in the direction of
\this\ is in the velocity range 30--40~km/s as shown in
~\ref{fig::Nanten}. The size of this cloud is $\sim70$~pc $\times
40$~pc in Galactic longitude and latitude (at an assumed distance of
4~kpc) and its total mass is estimated to be $2.5 \times 10^5$ solar
masses from the $^{12}$CO(J=1--0) velocity integrated intensity,
equivalent to 10$^{21}$ cm$^{-2}$ in molecular column density,
assuming an $X$-factor of $2 \times 10^{20} \mathrm{cm}^{-2} /
(\mathrm{K} \mathrm{km} \mathrm{s}^{-1})$~\citep{Bertsch} (where $X =
N(\mathrm{H}_{2})/I(\mathrm{CO})$, is the conversion factor between
H$_{2}$ column density and integrated CO intensity). This mass is
typical for a giant molecular cloud. \this\ is clearly located outside
of the dense cloud on a scale of $\sim$20~pc and therefore the cloud
and the VHE $\gamma$-ray emission are probably unrelated. The region
surrounding \this\ is a fairly complicated ensemble of young massive
stars as indicated by several compact HII
regions. Figure~\ref{fig::Spitzer} shows {\emph{Spitzer}} Space
telescope data of the GLIMPSE survey at 8~$\mu$m~\citep{Spitzer}. As
previously mentioned, no dust emission at the position of \this can be
found (red contours denote the 1, 2, and 3 $\sigma$ error contours on
the centroid of the VHE $\gamma$-ray emission), however, the peak in
the CO-data seems to be correlated to the 8~$\mu$m data, supporting a
star-formatting origin of the giant molecular cloud. The fact that
still a reasonable amount of dust in the region that has not yet been
blown away by the hot stars might suggests that these stars are still
rather young. The total integrated column density determined for a
beam size of 2.6\arcmin centred on Galactic longitude $12.8$, latitude
$0.0$ (the closest position to HESS\,J1813--178 in the NANTEN survey
with a grid size of 4\arcmin) is $\sim 9 \times 10^{22}
\mathrm{cm}^{-2}$, somewhat higher than the value determined
by~\citet{DickeyLockman} ($1.6 \times 10^{22}$ cm$^{-2}$), but fully
consistent with the high absorption density determined from the
{\emph{XMM-Newton}} data.  Integrating the column density only to
4~kpc, the proposed location of AX\,J1813--178 results in a value of
$\sim (0.5-1.) \times 10^{22} \mathrm{cm}^{-2}$. Comparing the NANTEN
values presented here with the values from~\citet{DickeyLockman} the
differences are due to both the difference in material probed (atomic
versus molecular hydrogen) as well as the angular resolution of the
two surveys.

\section{Interpretation of the X-ray emission, connection to other
wavebands}

Even though the NANTEN CO-Emission is probably unrelated to \this, it
is interesting to calculate an upper limit on the cosmic ray density
in the dense molecular cloud given the non-detection of a VHE
$\gamma$-ray signal. To calculate the ratio $k$ of the cosmic ray
density in the cloud $\rho_{\mathrm{CR, GMC}}$ to the local cosmic ray
density $\rho_{\mathrm{CR, local}}$ the following relation is used: 
\begin{equation}
k= \rho_{\mathrm{CR, GMC}} / \rho_{\mathrm{CR, local}} =
\frac{F_{(> 1~\mathrm{TeV})} \times D^2} {2.85 \times 10^{-13} \times M_5
\quad (\mathrm{kpc}^2 \ \mathrm{cm}^{-2}/\mathrm{s})}
\end{equation}
(here $F_{(> 1~\mathrm{TeV})}$ is the $\gamma$-ray flux above 1~TeV in
units of cm$^{-2}$ s$^{-1}$, $D$ is the distance in kpc, and $M_5$ is
the mass of the cloud in units of $10^5$ solar
masses)~\citet{FelixMoC}. The factor $2.85 \times 10^{-13}
\mathrm{kpc}^2 \ \mathrm{cm}^{-2}/\mathrm{s}$ has been derived
assuming a CR spectrum with a photon index of $2.6$.  Using a $2
\sigma$ H.E.S.S.\ upper limit of 0.6\% of the Crab flux above 1~TeV
(taken from the 10~hours exposure time and the H.E.S.S.\ sensitivity
of 1\% Crab flux in a 25~hours observation), i.e. $1.1 \times
10^{-13}$ cm$^-2$ s$^{-1}$, a distance of $D=4$~kpc, and a cloud
density of $M_{5} = 2.5$, an upper limit on $\rho_{\mathrm{CR, GMC}}
\sim 3 \times \rho_{\mathrm{CR, local}}$ can be derived.

Figure~\ref{fig::connection} shows a comparison between the X-ray and
Radio emission in the region surrounding \this. As can be clearly
seen, the radio emission shows a shell-like structure whereas the
X-ray emission has a compact core with an extended emission towards
the north-east, a typical morphology for a PWN~(see
\citet{GaenslerReview} for a recent review). The apparent
anti-correlation with the radio shell also suggest a confinement of
the X-ray emission within the shell, especially since the tail of the
X-ray source extends to the NE, where a break in the radio shell is
present. The non-detection of a pulsar in reasonably deep radio
observations may be due to beaming effects. In the X-ray data
presented here the thermal emission from the possible neutron star
might be buried underneath the strong non-thermal emission and
additionally strongly absorbed by the high column density which
strongly affects the detection of X-rays below $\sim
1.5$~keV. However, to finally confirm this scenario, the pulsar within
this nebula would need to be found, either in deep radio or X-ray
observations. Alternative strong evidence that AX\,J1813--178 is a PWN
would be the detection of spectral softening away from the core, a
signature of electron cooling which is observed in many PWN, see for
example \object{G21.5--0.9} \citet{SlaneG21.5}. The
{\emph{XMM-Newton}} data do not allow such an effect to be resolved in
AX\,J1813--178. A search for a steepening in the data with respect to
the distance of the best fit centroid yielded no conclusive results
within statistical errors. Nevertheless, it seems very likely that
AX\,J1813--178 is indeed a PWN, as the positional coincidence of a
hard spectrum X-ray source within a radio shell of G12.82--0.02 is
otherwise very unlikely. Furthermore, a chance positional coincidence
of the VHE $\gamma$-ray source \this\ with this composite radio/X-ray
object also seems unlikely, an in the following discussed we will
assume that the emission seen in these three wavebands originates in a
single new composite SNR, similar in its properties to e.g.\
\object{G\,0.9+0.1}~\citep{HESSG09}. However, the situation in the
case discussed here is somewhat different, since the VHE $\gamma$-ray
source shows extended emission (possibly owing to the smaller distance
of \this\ ($\sim 4$ kpc) in comparison to G\,0.9+0.1 ($\sim 8$ kpc)).

Both the shell of the SNR as well as the newly found PWN candidate
AX\,J1813--178 are viable $\gamma$-ray emitting objects. To estimate
whether the shell of the SNR could be responsible for the $\gamma$-ray
emission the following prescription has been followed: a background
(determined in a ring of 2--4\arcmin\ around the SNR centre) has been
subtracted from the central 2\arcmin\ radio emission, the resulting
radio map has been smoothed with a Gaussian kernel of width
$r_{\mathrm{smooth}} = 1.2\arcmin$, slices have been fit through the
resulting smoothed map in RA- and Dec-Direction with Gaussians to
determine $\sigma_x$ and $\sigma_y$ of the smoothed emission
region. Finally the ``Gaussian equivalent width'' of the SNR has been
defined as $\sigma_{\mathrm{SNR}} = (\sigma_x^2 + \sigma_y^2 -
r_{\mathrm{smooth}}^{2})^{1/2}$, yielding
$\sigma_{\mathrm{SNR}}\approx 1.8$\arcmin, compatible with the
measured rms extension $2.16\arcmin\ \pm 0.36\arcmin$ of the VHE
$\gamma$-ray emission region. The SNR shell seems therefore to be a
plausible candidate for the origin of the $\gamma$-ray emission.

As a result, we have to conclude that even with high angular
resolution {\emph{XMM-Newton}} X-ray observations in which the central
X-ray emitting object AX\,J1813--178 was resolved, no final
distinction can be drawn between the scenario in which the
$\gamma$-rays originate in the shell or in the core of G12.82--0.02.
The X-ray data described here and previously described radio
observations indicate that G\,12.82--0.02 is a composite supernova
remnant with a bright X-ray core and a radio shell. The size of the
gamma-ray source measured by H.E.S.S.\ appears to be consistent with
an origin of high energy emission in the SNR shell, but a common
origin of the X-ray and $\gamma$-ray emission in a central PWN cannot
be excluded as a larger spatial extent of the $\gamma$-ray source with
respect to the X-ray source could occur in such cases (see for example
the case of HESS\,J1825--137~\citep{HESSJ1825II}), due to the energy
dependent cooling of electrons.  Two distinct scenarios for the origin
of the TeV emission must therefore me considered: 1) as a counterpart
to the X-ray emitting core, and 2) as a counterpart to the radio
emission of the shell.

\begin{figure*}
  \centering
  \includegraphics[width=0.9\textwidth]{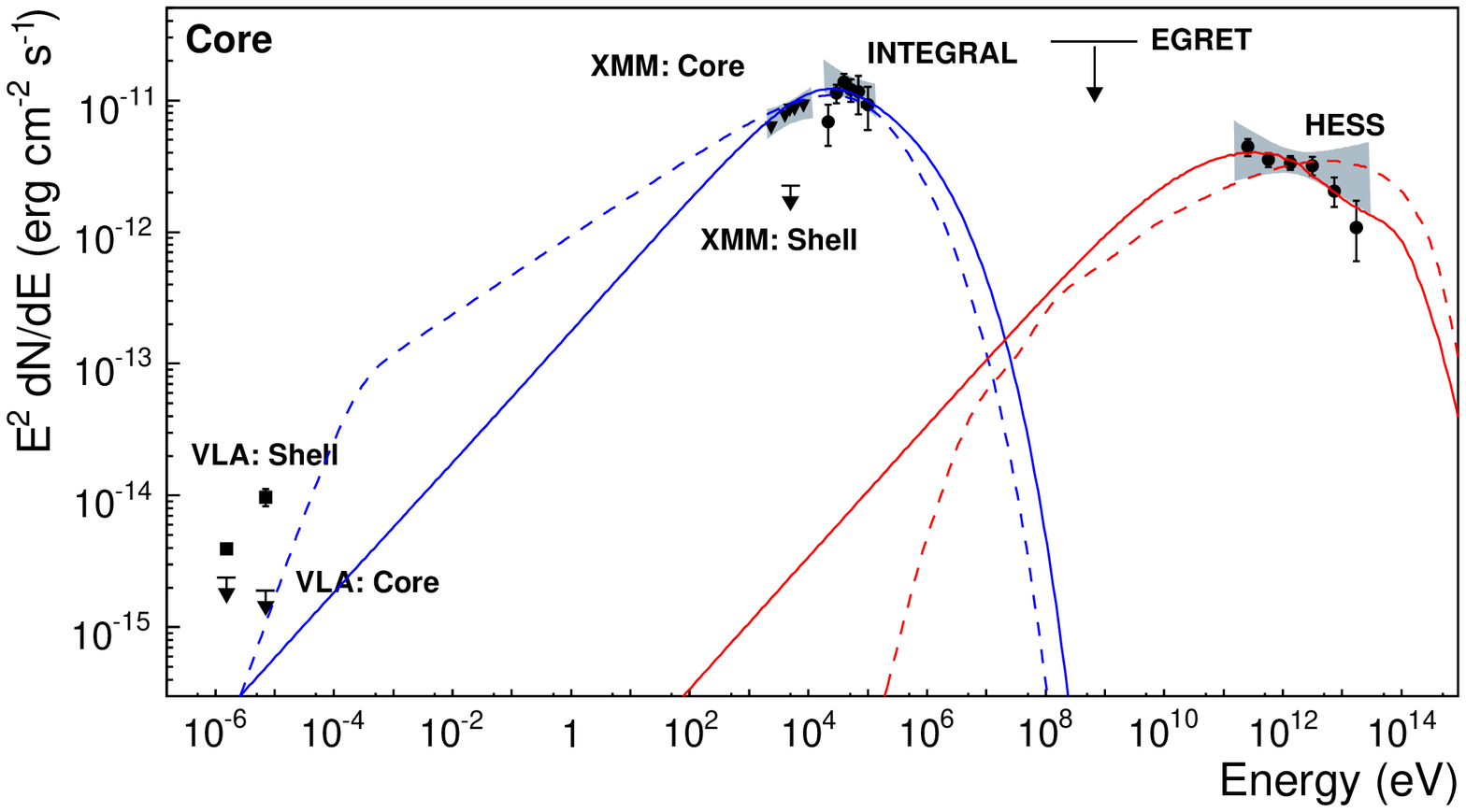}
  \includegraphics[width=0.9\textwidth]{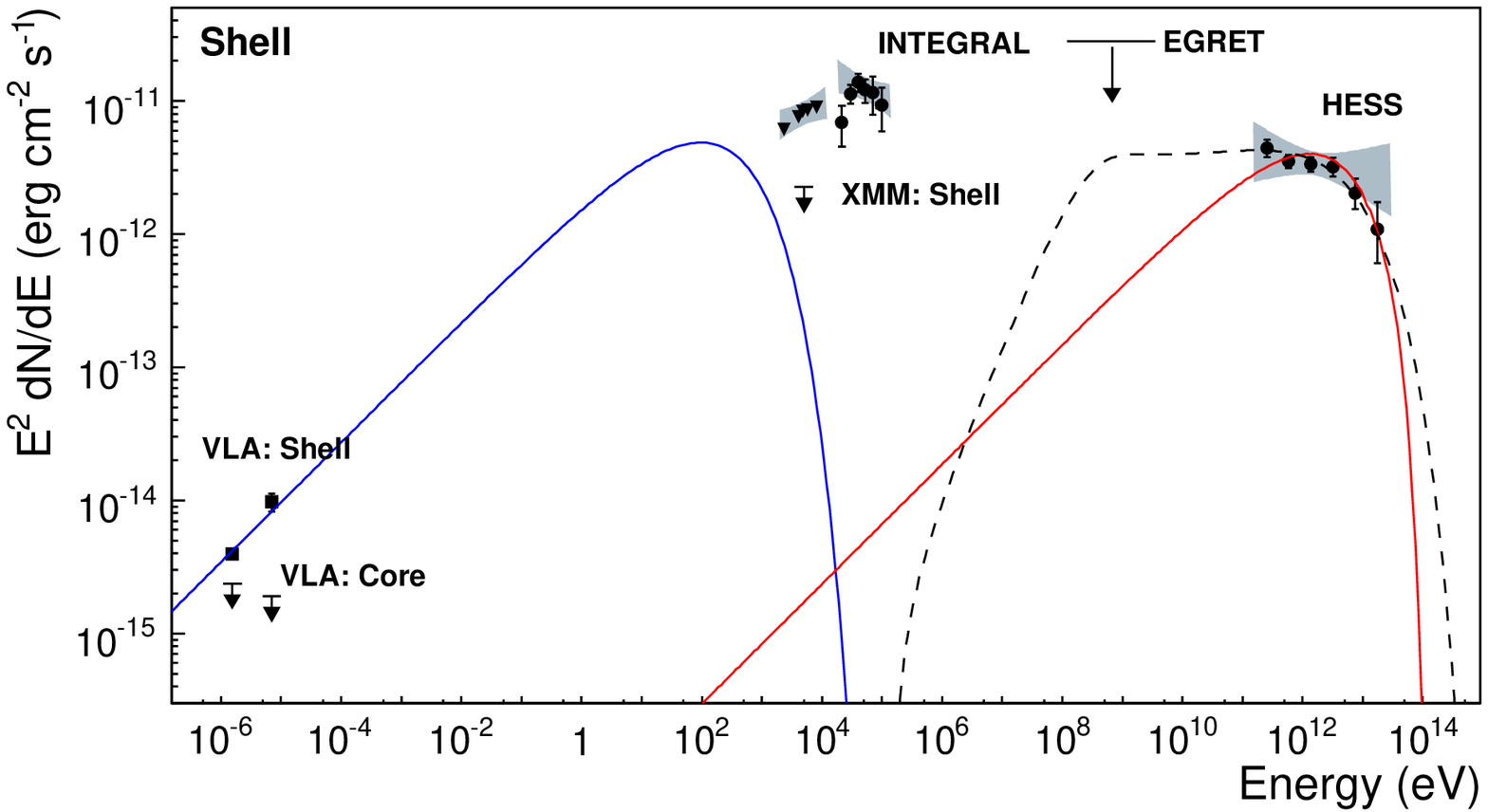}
  \caption{Spectral energy distribution for \this. The
  {\emph{XMM-Newton}} data 2~keV and 10~keV are shown for an
  extraction radius of 75\arcsec\ and are corrected for the
  absorption. A systematic error band of 0.2 on the spectral index and
  20\% on the flux level has been added. For the INTEGRAL points the
  data published in~\citet{Ubertini} have been recently reanalysed to
  determine error bars on the flux points. The flux points shown here
  are consistent with the published points. The H.E.S.S.\ spectrum has
  been rebinned in comparison to the energy spectrum shown
  in~\citet{HESSScanII}. {\bf{Top:}} Scenario in which the VHE
  $\gamma$-ray emission originates in the core of the SNR (i.e. the
  PWN). Leptonic one-zone model~\citep{AhaAto99} for two scenarios in
  which the VHE $\gamma$-ray emitting and the X-ray emitting electrons
  belong to the same population. The model parameters are specified in
  the text. {\bf{Bottom:}} Scenario in which the VHE $\gamma$-ray
  emission originates in the radio shell of the SNR. Leptonic (solid)
  and hadronic (dashed) VHE $\gamma$-ray emission scenarios in which
  the $\gamma$-ray emission originates in the shell of the SNR,
  i.e. the radio and VHE $\gamma$-ray emitting electrons are
  connected.}
  \label{fig::1813SED}
\end{figure*}

Scenario 1) has previously been discussed by~\citet{Ubertini}
and~\citet{Helfand}. Here the situation is revisited in the light of
the new {\emph{XMM-Newton}} data.  Figure~\ref{fig::1813SED}~(top)
shows the spectral energy distribution of G\,12.82--0.02. The EGRET
upper limit has been derived from the first 5~years of the EGRET
mission yielding a flux upper limit of $2.7 \times 10^{-11}$~erg
cm$^{-2}$ s$^{-1}$ above 100~MeV. Where angular resolution is
sufficient (i.e. in the radio and $<15$~keV X-ray bands) the core and
shell of the remnant are shown separately. Two synchrotron/Inverse
Compton model curves are shown for a population of relativistic
electrons in the core. The model was chosen to be time-dependent with
constant injection over the lifetime of the source. The key model
parameters are the slope of the injection spectrum of electrons
$\alpha$ the minimum and maximum energies of the electrons
$E_{\mathrm{min}}$ and $E_{\mathrm{max}}$, the magnetic field in the
source $B$, and the target radiation field (CMBR, optical or dust
photons).  In both cases shown an age of 1000 years is adopted,
in-line with the estimates of~\citet{Brogan} for the age of the
remnant: 300--3000 years. The combined {\emph{XMM-Newton}}/INTEGRAL
spectrum indicates that $E_{\mathrm{max}}$ must be rather high:
$\sim\,10^{15}$ eV. If inverse Compton emission takes place in the
Thompson regime then equal keV synchrotron X-ray and VHE $\gamma$-ray
IC spectral indices are expected away from any cut-off in the electron
spectrum. The softer spectrum measured by H.E.S.S.\ therefore suggests
that Klein-Nishina (KN) effects may be important in this source. The
well-detected X-ray peak determines roughly a convolution of the
magnetic field with the square of the maximum electron energy $B
\times E_{\mathrm{max}^2}$, whereas the ratio of the total synchrotron
to IC emission determines the ratio of $B^2$ to the radiation density,
i.e. the magnetic field $B$ if assuming normal radiation
fields. Therefore the electron index $\alpha$ is the only free
parameter in the modelling when assuming normal radiation fields. A
rather soft electron spectrum is needed to match the $\gamma$-ray
data, whereas the X-ray data suggests a harder spectrum. The dashed
lines in Figure~\ref{fig::1813SED}~(top) show a model with
$\alpha=2.4$, $B=4.2~\mu$G, $E_{\mathrm{min}}=25$~GeV,
$E_{\mathrm{max}}= 1.5$ PeV and a radiation field with near and far
infra-red components each of energy density 1 eV cm$^{-3}$ in addition
to the CMBR. Such a scenario is in marginal agreement with the
H.E.S.S.\ and X-ray data, but a low energy break or cut-off in the
electron spectrum is required to avoid over-producing radio emission
in the SNR core. This cutoff could be explained by the termination
shock of a PWN, in which the electrons in the lab-system have gained a
minimum energy through bulk motion. This minimum energy can be as high
as $\sim 100$~GeV up to 1~TeV, for a typical PWN Lorentz factor
between $10^{5}$ and $10^6$. The spectral break can be avoided by
invoking a strong contribution of scattering by IR/optical radiation
fields, where IC scattering transits into the KN regime and is thus
cut off at higher energies, resulting in an apparent steepening of the
VHE $\gamma$-ray spectra compared to the X-ray spectra.  The solid
curves in Figure~\ref{fig::1813SED}~(top) show a model with
$\alpha=2$, $B=7.5~\mu$G, $E_{\mathrm{min}}=1$~MeV, $E_{\mathrm{max}}=
1.5$~PeV and a radiation field with a very strong NIR component, with
energy density 1000 eV cm$^{-3}$.  This radiation field exceeds by a
factor of $\approx$ 1000 the nominal NIR radiation field at 4 kpc from
the Galactic Centre (see e.g.~\citet{PorterMoskalenko}) and is thus
somewhat unrealistic.  The (possibly) nearby star forming region W\,33
may contribute to the radiation density in the vicinity of the
source. However,~\citet{Helfand} have estimated the radiation density
in G\,12.82--0.02 to be 3--4 eV cm$^{-3}$, including the contribution
of W\,33.  Scenarios with intermediate values provide equal acceptable
agreement with the available data. See \citet{Hinton} for details of
the calculation methods used here.

A leptonic model, in which the core X-ray and VHE $\gamma$-ray
emission are associated yields an unavoidable $E_{\mathrm{max}} >
1$~PeV, suggesting that HESS\,J1813--178 is a highly effective
accelerator -- a Galactic {\emph{Pevatron}}. In case of a soft
electron injection spectrum $E_{\mathrm{max}}$ could well be higher,
since the {\emph{XMM-Newton}}--INTEGRAL spectral break could plausibly
be a cooling break, rather than represent the end of the electron
spectrum, for a somewhat greater pulsar age.  If the X-ray and VHE
$\gamma$-ray emission are indeed associated (i.e. originate from the
same electron population), the PeV maximum energy for the accelerated
electrons implies the emission almost certainly has plerionic
origin. PWN are certainly capable of accelerating electrons to PeV
energies as e.g.\ shown by EGRET measurements of the Crab Nebula,
showing a synchrotron component extending to MeV energies (shell type
SNRs in the framework of diffusive shock acceleration scheme fall
short of 1~PeV for \emph{electron} acceleration). The total energy
injected in the electrons is $6 \times 10^{46}$ ergs for the high-IR
case and even $6 \times 10^{47}$ ergs for the soft electron spectrum
case. For a source of age 1000 years (as suggested by~\citet{Brogan})
these values result in an electron luminosity of $L_{e} \sim 2 \times
10^{36} - 2 \times 10^{37}$ ergs/s. This value can be compared to the
comparable composite SNR G\,0.9+0.1 which has a derived electron
luminosity of $L_{e} \sim 7 \times 10^{37}$ erg/s~\citep{Hinton}.

A distinct alternative scenario is an origin of the $\gamma$-ray
emission in the SNR shell. Two young shell-type SNRs (with estimated
ages comparable to that of G\,12.82--0.02) are well established VHE
$\gamma$-ray sources, namely RX\,J1713.7--3946 and RX\,J0852.0--4622
(known as \emph{Vela Junior}).  Placed at the greater distance of
4~kpc, the TeV luminosity of HESS\,J1813--178 is comparable to that of
these two SNRs. Furthermore, the measured size of the TeV emitting
region is consistent with an origin in the radio shell.  The SNR shell
must therefore by considered seriously as an alternative source of the
TeV emission. Figure~\ref{fig::1813SED}~(bottom) shows two models for
a $\gamma$-ray origin in the SNR shell: a) as inverse Compton emission
from the electron population responsible for the radio emission (solid
line), and b) as the product of the decay of neutral pions produced in
hadronic interactions of accelerated protons in or near the SNR shell
(dashed line), calculated using the parametrisation of
\citet{Kelner}. In either case, the central X-ray emission has to
attributed to a different mechanism, presumably a pulsar. The electron
model has parameters: $\alpha=2.1$, $B=5~\mu$G, $E_{\mathrm{min}}=1$
MeV, $E_{\mathrm{max}}= 30$~TeV and a nominal 4~kpc radiation field
(note that larger values of $E_{\mathrm{min}}$ up to 1~GeV are still
compatible with the radio emission). The small value of
$E_{\mathrm{max}} = 30$~TeV is required to avoid producing significant
X-ray synchrotron emission in the shell, for values much larger than
this, 2~keV emission from the shell should have been seen in this
{\emph{XMM-Newton}} observation. The proton model has $\alpha=2.1$,
$E_{\mathrm{min}}=1$~GeV, $E_{\mathrm{max}}= 100$~TeV.  To match the
TeV flux level in the p-p interaction scenario the product of the
total energy in protons and the ambient density must be:
$(E_{p}/10^{50}\,\mathrm{erg}) (n/1~\mathrm{cm}^{-3}) \approx 1$. For
likely densities greater than $1~\mathrm{cm}^{-3}$, the required
acceleration efficiency is $<10$\% for a typical SNR explosion energy
of $10^{51}$ erg. In case of a very large local density,
Bremsstrahlung effects start to become important and might increase
the VHE $\gamma$-ray flux in comparison to the synchrotron X-ray
flux. However, as we consider most of the absorption in the X-ray
spectrum to be foreground density, the density within the source is
likely insufficient for Bremsstrahlung to be the dominant process.

\section{Summary and Conclusion}
Detailed {\emph{XMM-Newton}} X-ray and NANTEN $^{12}$CO(J=1--0)
studies have been performed towards the $\gamma$-ray source \this. The
NANTEN data show a giant molecular cloud in the vicinity of \this\
that might have played an important role in the evolution of the
$\gamma$-ray source. The X-ray data show a resolved non-thermal object
in the centre of the shell of the radio SNR, most likely representing
synchrotron emission from a PWN. The confirmation of this picture
would require the detection of a central pulsar, either in X-ray or in
radio observations, or indirectly through the detection of spectral
cooling as seen in several other PWN systems. Nevertheless,
G\,12.82--0.2 is now tentatively established as a {\emph{composite
SNR}} through its distinct morphology of a central extended
non-thermal object within a radio shell. For the first time, an object
of this type has been detected first in VHE $\gamma$-rays and then
identified with high-angular resolution radio and X-ray data. This
detection shows, that $\gamma$-ray observations are well suited to
identify SNR, that are otherwise very hard to detect due to
obscuration. No distinction is so far possible between a scenario in
which the $\gamma$-rays are emitted from the shell of the SNR, and one
in which they are emitted from the central PWN. An upcoming deep
Suzaku exposure on \this\ will shed more light on the situation in the
hard X-ray band and the future GLAST satellite will provide important
constraints in the MeV--GeV band. If the central X-ray source and the
VHE $\gamma$-ray source are indeed connected our modelling of the SED
emission suggests that \this\ is a Galactic Pevatron.

\begin{acknowledgements}
  The authors would like to acknowledge the support of their host
  institutions, and additionally support from the German Ministry for
  Education and Research (BMBF). Specifically, SF acknowledges support
  of the Department of energy (DOE). JAH is supported by a UK Particle
  Physics \& Astronomy Research Council (PPARC) Advanced Fellowship.
  We would like to thank the whole \HESS\ collaboration for their
  support. We would also like to thank Y.~Uchiyama for providing the
  analysis of the ASCA data. SF would like to thank C.~Brogan and
  B.~Gaensler for useful discussions on this source.
\end{acknowledgements}

\bibliographystyle{aa}

\begin{thebibliography}{}
\bibitem[Aharonian(1991)]{FelixMoC} Aharonian, F.~A., 1991, Ap\&SS 180,
  305  
\bibitem[Aharonian \& Atoyan (1999)]{AhaAto99}Aharonian, F.~A., \&
  Atoyan, A., 1999, A\&A 351, 330
\bibitem[Aharonian et al.(2005a)]{HESSScan} Aharonian, F.~A., et al. ({\it
  H.E.S.S.\ Collaboration}) 2005a, Science 307, 1938
\bibitem[Aharonian et al.(2005b)]{HESSLS5039} Aharonian, F.~A., et al. ({\it H.E.S.S.\
  Collaboration}), 2005b, Science 309, 746
\bibitem[Aharonian et al.(2005c)]{HESSG09} Aharonian, F.~A., et al. ({\it
  H.E.S.S.\ Collaboration}) 2005c, A\&A 432, 25
\bibitem[Aharonian et al.(2006a)]{HESSScanII} Aharonian, F.~A., et al. ({\it
  H.E.S.S.\ Collaboration}) 2006a, ApJ 636, 777
\bibitem[Aharonian et al.(2006b)]{HESSLS5039II} Aharonian, F.~A., et al. ({\it H.E.S.S.\
  Collaboration}), 2006b, A\&A in press
\bibitem[Aharonian et al.(2006c)]{HESSJ1825II} Aharonian, F.~A., et
  al. ({\it H.E.S.S.\ Collaboration}), 2006c, A\&A, in press
  (astro-ph/0607548) 
\bibitem[Albert et al.(2006)]{1813MAGIC}Albert, J., 2006, ApJ 673, L41
\bibitem[Benjamin et al.(2003)]{Spitzer}Benjamin, R.~A. et al., PASP
  115, 953
\bibitem[Bertsch et al.(1993)]{Bertsch} Bertsch, D.~L., et al., 1993, ApJ
  416, 587
\bibitem[Brand \& Blitz(1993)]{Brand} Brand, J. \& Blitz, L., 1993,
  A\&A, 275, 67
\bibitem[Brogan et al.(2005)]{Brogan} Brogan, C.~L., et al. 2005, ApJ
  629, L105
\bibitem[Churchwell(1990)]{UltracompactHII}Churchwell, E. 1990,
  A\&ARv, 2, 79
\bibitem[Dickey \& Lockman(1990)]{DickeyLockman} Dickey, J.~M., \&
  Lockman, F.~J., 1990, ARA\&A. 28, 215
\bibitem[Funk(2006)]{FunkBarcelona} Funk, S., 2006, to be
  published in Astrophys Space Science (astro-ph/0609586) 
\bibitem[Goldsmith \& Mao(1983)]{Goldsmith} Goldsmith, P.~F. \& Mao,
  X.-J., 1983, ApJ 265, 791
\bibitem[Gaensler \& Slane(2006)]{GaenslerReview}Gaensler, B.~M., \&
  Slane, P.~O., 2006, ARA\&A 44 (1), 17
\bibitem[Helfand et al.(2005)]{Helfand} Helfand, D.~J., Becker, R.~H.,
  \& White, R.~L., 2005, submitted to ApJL (astro-ph/0505392)
\bibitem[Hinton \& Aharonian(2006)]{Hinton}Hinton, J.~A., \& Aharonian,
  F.~A., 2006, ApJ, in press.
\bibitem[Kelner et al.(2006)]{Kelner}Kelner, S.~R., Aharonian, F.~A., \&
  Bugayov, V.~V., 2006, PhysRevD., 74 (3), 034018
\bibitem[Landi et al.(2006)]{Swift1813} Landi, R., et al., 2006, ApJ,
651, 190
\bibitem[Manchester et al.(2001)]{Parkes} Manchester, R.~N., 2001,
  MNRAS, 328, 17
\bibitem[Mitchell et al.(1990)]{Mitchell} Mitchell, G.~F., et al.,
  1990, ApJ 363, 554
\bibitem[Mizuno \& Fukui(2004)]{NantenCO} Mizuno, A., \& Fukui, Y.,
  2004,  ASP Conf. Ser. 317, p59
\bibitem[Porter et al.(2006)]{PorterMoskalenko} Porter, T.~A.,
  Moskalenko, I.~V., \& Strong, A.~W., 2006, ApJ 648 (2), L29
\bibitem[Reifenstein et al.(1970)]{Reifenstein} Reifenstein, E.~C., et
  al., 1970, A\&A 4, 357
\bibitem[Slane et al.(2000)]{SlaneG21.5}Slane, P., Yang, C., Schulz,
  N.~S., Seward, F.~D., Hughes, J.~P. \& Gaensler, B.~M. 2000, ApJ
  533, L29
\bibitem[Snowden, Collier \& Kuntz(2004)]{Snowden} Snowden, S.~L, Collier,
  M.~R, \& Kuntz, K.~D., 2004, ApJ 610, 1182
\bibitem[Sugizaki et al.(2001)]{ASCA}Sugizaki, M., Mitsuda, K., Kaneda,
  H., Matsuzaki, K., Yamauchi, S., \& Koyama, K. 2001, ApJS 134, 77
\bibitem[Ubertini et al.(2005)]{Ubertini} Ubertini, P. et al. 2005,
ApJ 629, L109
\bibitem[White et al.(2005)]{White} White, R.~L., Becker, R.~H., \&
  Helfand, D.~J., 2005, AJ 130, 586
\end{thebibliography}

\end{document}